\documentclass[12pt]{article}
\usepackage{amsmath}
\usepackage{graphicx,psfrag,epsf}
\usepackage{enumerate}
\usepackage{natbib}
\usepackage{textcomp}
\usepackage[hyphens]{url} 
\usepackage{hyperref}
\providecommand{\tightlist}{%
  \setlength{\itemsep}{0pt}\setlength{\parskip}{0pt}}

\newcommand{\blind}{0}

\usepackage{xcolor}
\xdefinecolor{blue}{rgb}{0.28,0.65,0.9}
\hypersetup{
    colorlinks,
    linkcolor={blue},
    citecolor={blue},
    urlcolor={blue}
}

\usepackage{booktabs}

\usepackage{amssymb}
\usepackage[T1]{fontenc}

\newcommand{\Beckman}[1]{Beckman}
\newcommand{\Tackett}[1]{Tackett}
\newcommand{\Rundel}[1]{Rundel}
\newcommand{\Sullivan}[1]{Sullivan}

\newcommand{\BeckmanBlind}[1]{Inst.\ A}
\newcommand{\TackettBlind}[1]{Inst.\ B}
\newcommand{\RundelBlind}[1]{Inst.\ C}
\newcommand{\SullivanBlind}[1]{Inst.\ D}
\usepackage{booktabs}
\usepackage{longtable}
\usepackage{array}
\usepackage{multirow}
\usepackage{wrapfig}
\usepackage{float}
\usepackage{colortbl}
\usepackage{pdflscape}
\usepackage{tabu}
\usepackage{threeparttable}
\usepackage{threeparttablex}
\usepackage[normalem]{ulem}
\usepackage{makecell}
\usepackage{xcolor}

\begin{document}

\def\spacingset#1{\renewcommand{\baselinestretch}%
{#1}\small\normalsize} \spacingset{1}


\if0\blind
{
  \title{\bf Integrating computing in the statistics and data science
curriculum: Creative structures, novel skills and habits, and ways to
teach computational thinking}

  \author{
        Nicholas J. Horton \\
    Amherst College\\
     and \\     Johanna S. Hardin \\
    Pomona College\\
      }
  \maketitle
} \fi

\if1\blind
{
  \bigskip
  \bigskip
  \bigskip
  \begin{center}
    {\LARGE\bf Integrating computing in the statistics and data science
curriculum: Creative structures, novel skills and habits, and ways to
teach computational thinking}
  \end{center}
  \medskip
} \fi

\bigskip
\begin{abstract}
Nolan and Temple Lang (2010) argued for the fundamental role of
computing in the statistics curriculum. In the intervening decade the
statistics education community has acknowledged that computational
skills are as important to statistics and data science practice as
mathematics. There remains a notable gap, however, between our
intentions and our actions. In this special issue of the \emph{Journal
of Statistics and Data Science Education} we have assembled a collection
of papers that (1) suggest creative structures to integrate computing,
(2) describe novel data science skills and habits, and (3) propose ways
to teach computational thinking. We believe that it is critical for the
community to redouble our efforts to embrace sophisticated computing in
the statistics and data science curriculum. We hope that these papers
provide useful guidance for the community to move these efforts forward.
\end{abstract}

\noindent%
\vfill

\newpage
\spacingset{1.45} 

\hypertarget{introduction}{%
\section*{Introduction}\label{introduction}}
\addcontentsline{toc}{section}{Introduction}

In their 2010 paper ``Computing in the Statistics Curriculum'', Deborah
Nolan and Duncan Temple Lang noted that ``computational literacy and
programming are as fundamental to statistical practice and research as
mathematics'' and that ``these changes necessitate re-evaluation of the
training and education practices in statistics''
\citep{nolan_templelang_2010}. We couldn't agree more about the
fundamental role of computing and the need for change at all educational
levels. Over the last decade we've seen the role of computing in the
statistics curriculum change and grow. The tools have become better,
computing is now more established in almost every classroom, and
arguably most importantly, the development and success of modern
statistics has been enhanced by ideas of computational thinking.

Before introducing the articles in this special issue, we reflect on the
questions originally posed by Nolan and Temple Lang:

\begin{enumerate}
\def\labelenumi{\arabic{enumi}.}
\tightlist
\item
  When they graduate, what ought our students be able to do
  computationally, and are we preparing them adequately in this regard?
\item
  Do we provide students the essential skills needed to engage in
  statistical problem solving and keep abreast of new technologies as
  they evolve?
\item
  Do our students build the confidence needed to overcome computational
  challenges to, for example, reliably design and run a synthetic
  experiment or carry out a comprehensive data analysis?
\item
  Overall, are we doing a good job preparing students who are ready to
  engage in and succeed at statistical inquiry?
\end{enumerate}

Nolan and Temple Lang also provided a damning critique of the status quo
at the time:

\begin{quote}
Many statisticians advocate-or at least practice-the approach in which
students are told to learn how to program by themselves, from each
other, or from their teaching assistant in a two-week ``crash course''
in basic syntax at the start of a course. Let us reflect on how
effective this approach has been. Can our students compute confidently,
reliably, and efficiently? We find that this do-it-yourself `lite'
approach sends a strong signal that the material is not of intellectual
importance relative to the material covered in lectures. In addition,
students pick up bad habits, misunderstandings, and, more importantly,
the wrong concepts. They learn just enough to get what they need done,
but they do not learn the simple ways to do things nor take the time to
abstract what they have learned and assimilate these generalities. Their
initial knowledge shapes the way they think in the future and typically
severely limits them, making some tasks impossible. (page 100)
\end{quote}

We concur that such an approach to computation is insufficient and at
times counterproductive.

What has happened in the intervening decade? We believe that there is a
growing consensus on the importance of computational literacy and
computing in the statistics and data science curriculum. The American
Statistical Associations updated Guidelines for Undergraduate Programs
in Statistics \citep{asa_guidelines_2014}, the revised GAISE (Guidelines
for Assessment and Instruction in Statistics Education) College report
\citep{gaisecollege}, and the National Academies of Science,
Engineering, and Medicine's consensus study on ``Data Science for
Undergraduates: Opportunities and Options'' \citep{nasem_2018} provide
detailed rationales for the fundamental role computing plays in
statistical thinking.

More pointedly, George \citet{cobb_2015} noted a convergence of
mathematics, computation, and context in statistics education and called
for a deep-rethinking of the curriculum from the ground up. The ``Mere
Renovation is Too Little Too Late'' paper sparked 19 spirited responses
and a provocative rejoinder (more on the ``tear-down'' metaphor) that
challenged the community in a number of fundamental ways
\citep{cobb_2015_discuss}.

We envisioned this special issue as a way both to highlight innovations
and approaches that have helped move the profession forward, as well as
to identify places where future work is needed. Many of these papers
work to answer the questions posed by Nolan and Temple Lang as well as
ones they had not anticipated in 2010.

The set of articles included in the special issue can be organized into
three non-mutually exclusive clusters that take different approaches to
address the questions laid out by Nolan and Temple Lang. The first
approach features \textbf{creative structures} for changing how we
integrate computing into the learning of statistics. The second approach
focuses on \textbf{novel or technical data science skills and habits}.
The third reflects that, more and more, statistics educators are
embracing and teaching ideas of \textbf{computational thinking}.

\hypertarget{creative-structures}{%
\subsubsection*{Creative structures}\label{creative-structures}}
\addcontentsline{toc}{subsubsection}{Creative structures}

Restructuring how we conceive of a syllabus and how we teach particular
material is never a small task. However, as different individuals
modernize their own courses, we can all learn from their experiences.
Both \citet{mine} and \citet{ellis} describe creative and modern data
science courses that fold together aspects of statistical inference with
vital computational skills. \citet{schwab-mccoy} report on a study
describing the emerging consensus of the elements of a data science
course. \citet{kim} present some of the technical aspects vital to
getting a solid computational course up and running. A less technical
approach is described by \citet{burckhardt} using the suite of materials
implemented by their Integrated Statistics Learning Environment (ISLE).
An immersive data science living and learning community is presented by
\citet{ward}. Finally, \citet{theobold} describe an alternative to
course learning through a series of workshops.

\hypertarget{novel-or-technical-data-science-skills-and-habits}{%
\subsubsection*{Novel or technical data science skills and
habits}\label{novel-or-technical-data-science-skills-and-habits}}
\addcontentsline{toc}{subsubsection}{Novel or technical data science
skills and habits}

The world of data science is rapidly changing, and it can be incredibly
difficult to keep up. Many of the papers in this special issue focus on
new, important, and exciting skills and tools that are important for
students if they want to contribute in today's data-centric world.
\citet{boehm} and \citet{mine} discuss the full cycle of iterating a
data science project. \citet{kim-hardin} take it one step further and
describe the importance of iterating on the full cycle. A few specific
skills are laid out in detail: \citet{dogucu} describe web scraping and
\citet{cummiskey} explore techniques for working with multivariate data.
\citet{horton} compare ways of incorporating Git in the statistical
classroom so that students have the skills to hit the ground running in
jobs and in their own data projects, reinforcing the value of
reproducible workflows as a foundation for reproducible research.

\hypertarget{computational-thinking}{%
\subsubsection*{Computational thinking}\label{computational-thinking}}
\addcontentsline{toc}{subsubsection}{Computational thinking}

The last approach may be the most difficult for statistics and data
science educators to embrace and implement in their own classes. The
value of bringing in ideas of software engineering or computational
thinking is that they help create a mindset that empowers students to
simultaneously think both statistically and computationally.
\citet{wing} describes how computing can impact a field, for example,
``Computer science's contribution to biology goes beyond the ability to
search through vast amounts of sequence data looking for patterns. The
hope is that data structures and algorithms---our computational
abstractions and methods---can represent the structure of proteins in
ways that elucidate their function. Computational biology is changing
the way biologists think.''

As a discipline, we are embracing the many ways that computing is
changing how statisticians think. \citet{woodard} report on a study
where students spoke through their thought process as they performed
computational tasks; their results are, somewhat unsurprisingly, that
computing is difficult and not intuitive. \citet{schwab-mccoy} describe
the challenge in front of us to teach computational thinking
effectively. \citet{ellis}, who describe debugging, and
\citet{theobold}, who discuss the teaching of iteration (a fundamental
component of algorithmic thinking), speak to integrating small pieces of
computational thinking within the data science curriculum.
\citet{reinhart} describe an entire course that is a cross between
software engineering and statistics, providing insight into the types of
skills that many statisticians (at all levels) need for success.

\hypertarget{what-would-deb-and-duncan-say}{%
\subsubsection*{What would Deb and Duncan
say?}\label{what-would-deb-and-duncan-say}}
\addcontentsline{toc}{subsubsection}{What would Deb and Duncan say?}

As we worked on the special issue, we thought that there would be value
in asking the authors of \citet{nolan_templelang_2010} to share their
thoughts about the paper, what they saw as most valuable, and to peer
into the future. Their incisive and provocative retrospective leads off
the special issue \citep{nolan-tl-response}.

\hypertarget{conclusion}{%
\subsubsection*{Conclusion}\label{conclusion}}
\addcontentsline{toc}{subsubsection}{Conclusion}

The articles in the special issue encourage us to redouble our efforts
to embrace computing in the classroom, to constantly push ourselves to
learn more tools, and to let computational thinking make our own work
better. We believe that the leading thinkers of the next decade will be
those who seamlessly knit together tools from both statistics and
computing and that how we think about statistics will be informed by
complementary computational thinking. To forge ahead we need to
cultivate computing foundations throughout the statistics paradigm. It
is our hope that the papers in this special issue initiate a new way of
thinking for your and your students.

\pagebreak

\bibliographystyle{agsm}
\bibliography{bibliography.bib}

\end{document}